\documentstyle[12pt]{article}
\begin{document}

{\bf Assessing the risk from the depleted Uranium weapons used in Operation
Allied Force.}

\begin{center}
T.E.Liolios $^{\dagger}$$^{\ddagger}$\footnote{%
email:theoliol@physics.auth.gr}

{\footnotesize $^{\dagger}$Department of Theoretical Physics,University of
Thessaloniki,Thessaloniki 54006,Greece }

{\footnotesize $^{\ddagger }$Hellenic War College, BST 903, Greece }
\end{center}

{\bf Abstract}

The conflict in Yugoslavia has been a source of great concern due to the
radiological and toxic hazard posed by the alleged presence of depleted
uranium in NATO\ weapons. In the present study some worst-case scenaria are
assumed in order to assess the risk for Yugoslavia and its neighboring
countries . The risk is proved to be negligible for the neighboring
countries while for Yugoslavia itself evidence is given that any increase in
total long-term cancer mortality will be so low that it will remain
undetected. Local radioactive hotspots such as DU weapons fragments and
abandoned battle tanks, fortified or contaminated with DU, constitute a
post-war hazard which is not studied in this article.

\medskip

{\bf 1.Introduction}

Operation Allied Force (OAF) has been going on for weeks\footnote{%
The present work was submitted to Science \& Global Security during
Operation Allied Force. Refereed and accepted for publication} in Yugoslavia
, employing sophisticated weapons that carry the spectrum of radiological
contamination. Over the past decades there has been a tremendous effort in
weapons laboratories to use depleted uranium (DU) in conventional weapons in
order to enhance their penetrability or to strengthen armor panels (tanks,
artillery etc.). Depleted uranium is used in a number of armor-piercing
anti-tank munitions, such as those aboard American A-10 Warthog jets, Apache
helicopters, and M-1 Abrams and Bradley tanks. US. and Allied forces fired
approximately 315 tons of depleted uranium\cite{fasirak} during the Persian
Gulf War. Yugoslav state news media have referred to ''radioactive bombs''
being launched by NATO. There is strong likelihood that the weapons referred
to are composed of depleted uranium (DU). Its ability to self-sharpen as it
penetrates armor is the main reason why tungsten, which tends to mushroom
upon impact, has been abandoned. Nevertheless, the high temperatures caused
by the high explosives (HE) detonated in the weapon or the friction between
the ammunition and the target (armor, concrete....) lead to the generation
of uranium oxides which along with the tiny fragments of the weapon case
pose a serious radiological hazard to living beings. So far no measurement
has shown any increase in the environmental radioactivity either in
Yugoslavia or in Greece. As for Yugoslavia, one has to rely on the local
scientific community to detect and assess the contamination. However, as
there has been severe censorship on every sort of information by the Serbs,
and most likely by the NATO\ officials, the international scientific
community should independently attempt to reliably assess the possible
implications of DU that have allegedly been dropped in the Balkans. Until
some counter detects the contamination the only means available are
theoretical hazard predictions through computer simulations . By applying a
worst case scenario, an initial emergency assessment , or safety analysis
planning is possible. Although, precise data about the performance and the
composition of these weapons are classified, in a worst case scenario one
can use the available unclassified data which can still yield the magnitude
of the hazard and trigger an appropriate emergency planning and response. In
the present work a very reliable computer code has been used which simulates
explosions where nuclear material is involved. The code is ''HOTSPOT''\cite
{hotspot} produced by S.G.Homann at Lawrence Livermore National Laboratory .
It is a very effective Gaussian plume model suitable for radiation risk
assessments at short and large distances from the source. Moreover the
familiar wedge model\cite{wmod} is applied in order to predict average total
long-term cancer fatalities caused by DU inhalation. Throughout the present
work basic information is given about fundamental radiological properties or
weapons compositions. This is imperative as the results presented here are
expected to be of interest to non-experts, as well.

{\bf 2.NATO weapons overview in OAF.}

A thorough analysis of the weapons used by NATO\ against Serbia\cite{fasarms}
indicates that some of them are specially designed to penetrate hard targets
. Despite the fact that the precise data for those weapons are classified
there are some very strong arguments that indicate the presence of DU in
their composition.

a) Yugoslav state news media have referred to ''radioactive bombs'' being
launched by NATO \cite{msnbc}.

b) The Tomahawk currently in use is Tomahawk Block III with improved target
penetration \cite{gao}. The only material that can improve target
penetration nowadays is DU.

c) No data exist about the precise composition of the Tomahawk high density
penetrator which raises an eyebrow about the motives of this secrecy.

However, its warhead must be either a kinetic energy penetrator or a
multiple warhead system (MWS) \cite{hewish} with an approximate weight of $%
400\,kg\,$(plus $50\,kg$ of HE). In either case, penetration performance
depends on the weight per cross sectional area of the follow-through
warhead. That means that a very heavy metal has to be used in its
composition to maximize target penetration.

Some very penetrating weapons used in the current war are \cite{fasarms2}:

{\bf Tomahawk missile.} An all-weather submarine or ship launched
land-attack missile. It is used to attack a variety of hard fixed targets ,
which explains why the missile has to be extremely penetrating (which makes
it a DU suspect). During the war in the Gulf, 288 missiles where fired (II\
generation) while so far hundreds (III generation and probably some
experimental of the IV generation ) are believed to have struck targets in
Serbia and Kosovo. That highly sophisticated weapons carries a single
conventional warhead or submunitions. The BGM-109 model weighs 1192 kg, has
a length of 5.56 m, and a diameter of 51.8 cm ( without the booster ). A
rough estimate of the typical weight of its airframe is 400 kg \cite{tsipsa}$%
.$ Besides, in the same category we have to include the Air Force's
Conventional Air Launched Cruise Missile CALCM . CALCM used to carry nuclear
warheads and has been converted to conventional weapons. Its frame may have
been fortified with DU to withstand the blast of anti-missile defences of
the FUSSR. In any case, in a worst-case scenario, cruise missiles must be
considered DU carriers. The positive aspect of that weapon is that sample
fragments of its casing, scattered in the vicinity of the explosion, may
reveal its composition.

{\bf BLU-107 Durandal} .The Durandal anti-runway bomb was developed by the
French company MATRA, designed solely for the purpose of destroying runways.
Once the parachute retarded low-level drop bomb attains a nose-down
attitude, it fires a rocket booster that penetrates the runway surface, and
a delayed explosion buckles a portion of the runway. It can penetrate up to
40 centimeters of concrete, creating a 200 square meter crater causing
damage more difficult to repair than the crater of a general-purpose bomb.

{\bf BLU-109/B}. The BLU-109/B (I-2000) is an improved 2,000-pound-class
bomb designed as a penetrator without a forward fuze well. Its configuration
is relatively slim, and its skin is much harder than that of the standard
MK-84 bomb. The skin is a single-piece, forged warhead casing of one-inch,
high-grade steel. Its usual tail fuze is a mechanical-electrical FMU- 143.
The 1,925-pound bomb has a 550-pound tritonal high-explosive blast warhead

{\bf Guided Bomb Unit-28 (GBU-28)}. The Guided Bomb Unit-28 (GBU-28) is a
special weapon developed for penetrating hardened command centers located
deep underground. The GBU-28 is a 5,000-pound laser-guided conventional
munition that uses a 4,400-pound penetrating warhead. The bombs are modified
Army artillery tubes, weigh 4,637 pounds, and contain 630 pounds of high
explosives.

{\bf AGM-114 Hellfire II. } Laser Hellfire presently is used as the main
armament of the U.S. Army's AH-64 Apache and US. Marine Corps's AH-1W Super
Cobra helicopters. For antiarmor roles, the AGM-114 missile has a conical
shaped charge warhead with a copper liner cone that forms the jet that
provides armor penetration. This high explosive, antitank warhead is
effective against various types of armor including applique and reactive.
Actual penetration performance is classified.

{\bf The PGU-14/B API ammunition}. That Armor Piercing Incendiary round has
a lightweight body which contains a sub-calibre high density penetrator of
Depleted Uranium (DU). In addition to its penetrating capability DU is a
natural pyrophoric material which enhances the incendiary effects. It is
used by the AN/GAU-8 30mm Avenger ( a 30mm seven-barrel gatling gun, mounted
only on the A-10 attack jet, used primarily in the air to ground role as a
soft target killer and tank buster) and also by the M230 automatic gun
mounted on the Apache helicopter.

{\bf M256 120mm smoothbore cannon}. It is the main weapon of the M1A1 battle
tank. The primary armor-defeating ammunition of this weapon is the
armor-piercing, fin-stabilized, discarding sabot (APDS-FS) round, which
features a depleted uranium penetrator. Battle tanks have not been used yet
by the NATO\ forces, therefore that scenario is not studied for the time
being.

In our study we will focus our simulation on the Tomahawk missiles, the
BLU-109 bomb and the API ammunition, as not only do they represent well our
worst case scenaria but also the available unclassified information suffices
for our risk assessment approach. Note that the bomblet dispersion version
of Tomahawk is not expected to have an improved penetration capability and
therefore our models will focus on the single warhead version.

{\bf 3. A short description of DU}

Depleted uranium$\cite{du}$ is the metallic remnant of a series of processes
the uranium ore undergoes and it is roughly 60 percent as radioactive as
naturally occurring uranium. On the other hand, Uranium, a radioactive
element, is a silver-white metal in its pure form. It is a heavy metal
nearly twice as dense as lead $\left( 19\frac{gr}{cm^{3}}\right) $ compared
with $\left( 11.4\frac{gr}{cm^{3}}\right) $. On average, each of us takes in 
$1.9\,\mu g$ $\left( 0.65\times 10^{-6}\mu Ci\right) \,$of uranium a day
from food and water, and inhales a very small fraction $7\times 10^{-3}\mu g$
$\left( 2.3\times 10^{-9}\mu Ci\right) $ every day. In nature Uranium is
composed of three isotopes (each has its own unique decay process emitting
some form of ionizing radiation:alpha, beta, gamma radiation or a
combination) in the following ratio:

\begin{center}
{\bf NATURAL URANIUM COMPOSITION } 
\[
_{92}^{234}U\left( 0.0054\%\right) ,_{92}^{235}U\left( 0.7\%\right)
,_{92}^{238}U\left( 99.3\%\right) 
\]
\end{center}

In the gaseous diffusion process two fractions are produced in the form of $%
UF_{6}:$ one enriched in $^{235}U$ and the other depleted in $^{235}U.$ The
former is further processed to give weapons-grade Uranium (WgU) whereas the
latter is chemically transformed by weapons manufacturers into Uranium metal
and alloys, suitable for ammunition and armor panels.

In fact, DU has a low content of $^{234}U,$ and $^{235}U$ which have been
removed in the depletion process. Therefore the product and by-product of
the enrichment are respectively \cite{fetdetec}:

\begin{center}
{\bf WEAPON-GRADE URANIUM COMPOSITION}

\[
_{92}^{234}U\left( 1\%\right) ,_{92}^{235}U\left( 93.5\%\right)
,_{92}^{238}U\left( 5.5\%\right) 
\]

{\bf DEPLETED URANIUM COMPOSITION} 
\[
_{92}^{235}U\left( 0,2\%\right) ,_{92}^{238}U\left( 99,8\%\right) 
\]
\end{center}

After the enrichment process DU can used as a fusion tamper in the
thermonuclear weapons. The fusion tamper prevents the escape of thermal
radiation from the thermonuclear fuel thus enhancing the burn efficiency.
Moreover, fast neutrons ( $2.45 MeV$ and $14.1 MeV$) from the fusion
processes fission the DU tamper. This extra boost accounts for half the
yield of a fission-fusion-fission nuclear bomb \cite{nwfaq}.

The most important constituent of DU is $_{92}^{238}U$ , an alpha emitter
with a half-life of $4.5\times 10^{9}$ years and a specific activity of $%
3.4\times 10^{-7}\frac{Ci}{gr}$ (while the isotope $^{235}U$ has a specific
activity of $2.2\times 10^{-6}\frac{Ci}{gr}).$ The total combined specific
activity of DU is $4.76\times 10^{-7}\frac{Ci}{gr}\cite{hotspot2}$. It has
two short-lived daughters :($^{234}Th$, half-life of 24.1 days) and ($%
^{234}Pa$, half-life of 1.17 minutes) which are beta and weak gamma
emitters. Because of this constant nuclear decay process, very small amounts
of these ''daughters'' are always present in DU. On the other hand $^{235}U$
(half-life $7\times 10^{8}$ years) decays into $^{231}Pa$ (half-life $%
3.25\times 10^{4}$ years), which is an alpha, beta, and gamma ray emitter.
The $^{238}U$ and $^{235}U$ chains continue through a series of long-lived
isotopes before terminating in stable, non-radioactive lead isotopes $%
^{206}Pb$ and $^{207}Pb$. Note that regardless of its size (large fragments
or small particles), once entering the body, DU is subject to various
degrees of solubilization-it dissolves in bodily fluids, which act as a
solvent. Its main toxic effects are cellular necrosis and renal failure. The
American Conference of Governmental Industrial Hygienists (ACGIH) has
established a Threshold Limit Value (TLV ) \cite{tvl}of $0.2\frac{mg}{m^{3}}$%
\thinspace (for both soluble and insoluble compounds). TLVs are based on the
principle that there is a threshold below which no adverse health effects
occur and are called time-weighted-average values because they are averaged
over an 8-hour workday, for a 40-hour workweek over a working lifetime.
Though TVLs were developed for the working environment , in the battlefield
or in emergency planning they can still give a measure of the risk.

{\bf 4.DU cancer risk}

DU is radioactive and therefore carcinogenic. The principal hazard from
exposure to DU aerosols is an increased probability of cancer of the lung
and of other organs where the DU oxides are transported. While it is
difficult to calculate the total immediate radiation effects on health in
terms of exact doses to specific individuals, we can we resort to the
''wedge model'' \cite{wedge}in order to compute the average total long-term
man-rem doses. According to this model, the total amount $\left( I\right)
\,\,$of DU inhaled, as a result of a given release, is : $I=Qbpu^{-1},\,$%
where $Q$ the total amount of DU released, $b$ the breathing rate\cite{brate}%
, $p$ the average population density and $u$ the deposition velocity. If we
make the assumption that the risk is linear with dose then we can combine
the committed effective dose equivalent (CEDE )\cite{cede} for $^{238}U$ $\,$%
inhaled ($1.2\times 10^{8}\frac{rem}{Ci})$ with the ICRP cancer risk factor
( $5\times 10^{-4}\,\frac{cancers}{rem})\,$\cite{icrp} to estimate cancer
risks from DU inhalation with respect to population densities and deposition
velocities. The results are shown in Figure 1.

{\bf 5. Simulation of Tomahawk attacks .}

{\bf 5.1 The wedge model predictions}

In the present work we will limit our discussion in the conventional use of
DU as this is currently employed in Yugoslavia. It is common sense that most
of the attacks against industrial facilities , bridges and government
buildings need weapons with enhanced penetrability. That need spells the
name of DU. Such is also the case for anti-tank munition, anti-radar bombs
or weapons which destroy the runways of airports. The most infamous weapon
is the Tomahawk missile used day after day by the NATO alliance.

Being consistent with our worst case scenario we assume that the kinetic
energy warhead of the Tomahawk missile is made of $400\,kg$ of DU. Therefore$%
,$ we have an activity of $0.192\,Ci$ per missile. After the impact only a
small quantity will constitute the respirable fraction-defined as the
fraction of the released material associated with an Activity Median
Aerodynamic Diameter (AMAD) of $1\mu m.$ The default ICRP-30\cite{icrp}
internal dosimetry conversion factors also assume an AMAD particle-size
distribution of $1\mu m.$ During the explosion a temperature of $5000\,^{0}C$
is reached \cite{he} which exceeds the boiling point of Uranium ($%
4700\,^{0}C).\,\,\,$That temperature will produce a large quantity of DU
aerosols in the form of Uranium Oxides that may find their way into the
respiratory tract.

The previous analysis of the wedge model can be used to predict maximum
cancer lethality per missile. Assuming an attack in a densely populated
urban area (3000 km$^{-2})$, we have approximately one cancer per Tomahawk
missile. That is the total long-term lethality per missile assuming that all
the DU carried becomes respirable, which of course is a worst-case scenario
showing the maximum potential of a Tomahawk to cause cancer. (Note that the
media speak of several hundreds of such missiles fired against Yugoslavia
during Operation Allied Force).

{\bf 5.2. The Gaussian model predictions}

The default respirable fraction$\,$of $20\%$ in a uranium explosion, used in
the HOTSPOT, is indeed a realistic scenario to assess immediate risks,
particularly at short distances. That being the case, each Tomahawk is
supposed to carry a respirable radioactivity of $38\,\frac{mCi}{missile}.$
To realize the magnitude of that activity, a typical radioactive quantity
injected into a patient in a thyroid function test is $10\,\mu Ci$ that is
approximately $3800$ times less. On the other hand a typical amount of
radioactivity released in a large scale reactor accident is $10^{8}Ci,$ \cite
{fettsip}that is approximately $26\times 10^{8}$ more. The non-respirable
fraction which consists of fragments scattered in the vicinity of the
explosion, and particles much larger than $1\mu m$ will be ignored in the
rest of this study though the are highly toxic and will definitely be
localized and contaminate the vicinity of the explosion. Nor will we discuss
the aggravation of lethality due to open wound or injuries during the rescue
operations.

Of course during the explosion the distribution of the radioactive DU is
governed by such factors as wind speed, amount of explosives , deposition
velocity and so on that will further reduce the lethality of the missile.

In the model of this study we make the assumption that a single Tomahawk
strike is actually a $400$ $kg$ DU explosion which involves the detonation
of $50\,kg$ of HE. The release fraction is $20\%$ ( that is the percentage
of the warhead that can be inhaled after the explosion\cite{conf}) and the
wind speed is assumed to be $8$ $\frac{m}{\sec }$. The time of day is night
(stability class D) , while the deposition velocity is $1\frac{cm}{\sec }$.
Note that the cloud effective heights calculated by HOTSPOT agree well with
the experimental data for detonations of similar yields\cite{he2}.The
''HOTSPOT'' calculations yield the 50-years CEDEs (due to inhalation as the
ground shine is negligible compared to plume effects) and the ground
deposition of radioactivity at various distances (Figure 2., 3.). 
\[
{\small 
\begin{tabular}{llllllllllll}
Distance $\left( km\right) $ & 0.1 & 0.2 & 0.5 & 1 & 2 & 5 & 10 & 20 & 50 & 
75 & 100 \\ 
50-year CEDE $\left( mrem\right) $ & 4.8 & 3.9 & 2.6 & 1.7 & 1.0 & 0.5 & 0.26
& 0.13 & 0.046 & 0.029 & 0.021 \\ 
Ground depos.$\left( 10^{-3}\frac{\mu Ci}{m^{2}}\right) $ & 38 & 31 & 20 & 13
& 7.8 & 3.5 & 1.7 & 0.71 & 0.18 & 0.09 & 0.06
\end{tabular}
} 
\]

If we take into account that the current established protection standards
are:\cite{radlim}

a) 5 rems in a year for workers (to protect against cancer).

b) 50 rems in a year for workers to any organ (to protect against threshold

effects, such as radiation burns, etc.).

c) 50 rems in a year to the skin or to any extremity.

d) 15 rems in a year to the lens of the eye (to protect against cataracts).

e) 0.1 rem in a year (70-year lifetime) for members of the public.

we come to the conclusion that people who are as close as $100m$ at the time
of the explosion are expected to receive, over a period of 50 years after
the explosion, $20$ times less than the maximum allowed dose per year.
Needles to say, at distances larger than $20\,km$ the doses are negligible.
Of course, at close distances, the results of the blast wave will be
devastating and will prevail over any other effect.

The ground deposition, on the other hand, reaches the concentration of $0.038%
\frac{\mu Ci}{m^{2}}$ at a distance of $100m$ where we have to remember that
a concentration of $2\frac{\mu Ci}{m^{2}}$\thinspace is needed for land to
be rendered unsuitable for cultivation\cite{nwfaq2}, that is almost 50 times
more.

To underline the impossibility of DU radiological contamination for the
neighboring countries of Yugoslavia, we can assume that 1000 such attacks
are made against targets in Pristina in Northern Kosovo. That would cause a
50-CEDE of $0.046\,rem$ at a distance of $50$ km. Note that a CT exam
administers a dose of $1.1rem$ (head and body). If we rotate the isodose
downwind radius of the Gaussian model to cover all possible wind directions,
then a circular spot is produced which indicates the area at risk according
to the present model (Figure 4). Outside the borders of that circular area
the plume is not expected to deliver a dose higher than that of a pelvis
x-ray.

{\bf 6. Simulation of BLU-109/B bomb attacks.}

In that model , consistent with our worst case scenario, we also assume that
the warhead of the bomb is made of DU . In that case we have the explosion
of 651 kg of DU with 243 kg of HE. Therefore assuming the use of a quantity
of 1000 BLU-109/B against targets in Pristina and the same conditions as in
the Tomahawk case we obtain a 50-CEDE of $0.06\,rem\,$at a distance of 50
km. The combined CEDE of Tomahawks and BLUs would still be low: $0.1rem$
(less than a lumbar spinal x-ray).In fact, if such was the case then those
attacks would have dropped some $1000\,$tons ($200$ respirable tons) of DU
in Yugoslavia when according to the Iraqi authorities the war in the Gulf
left $315\,$tons of DU in Iraq.

{\bf 7. DU of the PGU-14/B API and the APDS-FS rounds .}

A typical combat load for the GAU-8 gun is 1,100 30 mm rounds. Each round
contains $330gr$ of DU , alloyed with $0.75\,$weight percent titanium. The
projectile is encased in $0.8\,mm$-thick aluminum shell as the final DU round%
\cite{rounds}, preventing any escape of the $a-$radiation emitted..
Consequently each round carries approximately $1.5\times 10^{-4}Ci.$Upon
impact, the shell is subject to high temperatures due to friction with the
armor panel. Moreover, if the armored vehicle explodes or is set on fire
then the respirable activity produced by the armor panel should also be
taken into account. For example, the Abrams battle tank's thicker armor is
reinforced at the turret and flanks by DU panels inserted between regular
steel armor. Another source of DU is the primary armor-defeating ammunition
of the M256 120mm smoothbore cannon (main weapon of the M1A1 battle tank),
which is an armor-piercing, fin-stabilized, discarding sabot round. It is
imperative that battle tanks, attacked by NATO forces in Yugoslavia, are
closely examined for radioactive traces. Note that the DU rounds always
leave a distinctive radioactive trace on the entrance and exit holes. Each
time an A10 unloads its gun, $360\,kg$ of DU will be released in the
environment. Assuming that an A10 unloads its gun on every mission and that
the whole DU quantity becomes respirable (worst case scenario) , the wedge
model can be used to predict total long-term cancer fatalities per mission
(Figure 5). As the average population of Yugoslavia is 100 km$^{-2}$ it
turns out that it would take roughly fifty A-10 missions to have an
additional cancer death. Although a commander would strive to deploy his
troops as sparsely as possible, in the theater of operations the average
population density of ground forces can reach urban area levels (300 km$%
^{-2}\,$to 3000 km$^{-2})$ . As a result in the battlefield cancer risk is
expected to be higher. If we assume for the people of Yugoslavia an
individual cancer death risk similar to that of the US. ($i.e.,20\%)$, then
5000 such attacks would increase individual cancer risk by $10^{-5}.$ That
increase would remain undetected against the large background. Of course, it
is very unlikely that such a number of attacks has occurred so far.

{\bf 8. The Hellfire case}. Due to its low yield and weight (warhead weighs
less than 10 kg) any risk associated with that weapon will be much less
significant than the Tomahawk scenario risk. If the classified composition
of its armor-piercing structure is indeed DU , then it is expected to pose a
hazard only to people at very close distances, especially during '' battle
tank fires''. Since no reliable data exist and no use of the weapon in
question has been made yet in Kosovo, any assumption might further
complicate the current situation.

{\bf 9. Chemical toxicity of DU.}

The toxic risk can be assessed by means of a simple model without knowing
details of the population over which it is dispersed and the meteorological
conditions. Suppose that $1000\,\,tons$ of respirable DU is dispersed
uniformly over Greece which has an area of $132.000\,km^{2}$ . We assume
that all the aerosols have been concentrated in a volume with $1km$ height.
This gives a concentration of $7.5\times 10^{-3}\frac{mg}{m^{3}}$, which is
about $26$ times less than the threshold limit value. A similar calculation
yields an air concentration of $\,0.04\frac{mg}{m^{3}}\,$for FYROM\cite
{fyrom} which should not cause much concern either.

The lifetime of the toxic cloud depends on the height and the rate at which
the particles fall out. A deposition velocity of $1\frac{cm}{\sec }$ is very
plausible\cite{hotspot3} while particles larger than $1\mu m\,$will fall
faster. Rain or moisture will increase that velocity. In that scenario,
particles from the top of the cloud will take 27 hours to reach the ground.
It is very unlikely that the cloud will remain over a city for that long.
Even a light breeze (5$\frac{m}{\sec })$ will carry the cloud beyond a large
city (the size of Athens) in a few hours.

Of course an actual toxic cloud is not expected to have the above shape but
the present model gives solid evidence that the fear of toxic poisoning, due
to DU that is allegedly used in the present war, is groundless. Note that
the amount of DU that could be inhaled is independent of the height and the
extend of the cloud as shown in a similar study that disproved exaggerated
allegations about Plutonium risks\cite{plut}.

That absolutely worst-case scenaria show that there is no immediate hazard
from the radiological or chemical toxicity of DU for the neighboring
countries of Yugoslavia. Admittedly, localized DU can enter the food chain
and reach inhabitants of other countries by means of exported goods or river
streams. However, such aspects are regarded as less harmful than actual
inhalation of the DU plume.

{\bf 10.Conclusions.}

We have assumed some worst-case scenaria in order to assess the radiological
and chemical risk of the alleged use of DU in OAF, for Yugoslavia and its
neighboring countries. For the time being, the risk for the neighboring
countries is found to be negligible , while for the people of Yugoslavia
itself, evidence is given that any increase in average total, long-term
cancer mortality will be so low that it will remain undetected.

The use of the PGU-14/B API ammunition seems to be the most hazardous weapon
used daily in the theater of operations as is openly declared a DU\ carrier.
Its use so far has been limited to the Avenger gun of the A-10 jet. If
Apache helicopters move in, the effects will escalate and need further
investigation. On the other hand, accurate data about the composition of the
weapons used in OAF are needed in order to accurately predict the
radiological and chemical contamination of DU at very close distances,
especially in order to investigate the formation of radioactive hotspots.
Such data could be either obtained by the NATO authorities or by studying
the fragments of the weapons in question (Tomahawk missiles, BLU bombs etc.)
scattered in the vicinity of the explosion. Once DU is detected, the above
simulations can be fed with more accurate data in order to perform a precise
risk assessment in the area.

{\bf ACKNOWLEDGMENTS}

The author would like to thank S.Homann for providing his code HOTSPOT, as
well as H.Feiveson, S.Fetter, K.Ypsilantis and A.Petkou for useful comments
and discussion.

The present work was inspired by an article which used the same risk
assessment models in order to study the hazards from Plutonium dispersal
accidents\cite{von}.

The author is grateful to C.Zerefos for valuable information and advice on
the post-war environmental hazards in Yugoslavia.


\begin{thebibliography}{99}
\bibitem{cede}  Commited Effective Dose Equivalent is a weighted sum of
organ dose equivalents multiplied by appropriate risk weighting factor. It
stands for the total dose a person is commited to receive as the result of
an intake of radioactive materials during the 50-year period after the
intake.
\bibitem{conf}  In fact an additional fraction of the actual plume is
expected to be confined inside the structure where the warhead explodes.
Therefore our scenario remains consistently a worst-case one.
\end{thebibliography}
\end{document}